\documentclass[sigconf]{acmart}

\usepackage{multirow}
\usepackage{graphicx}
\usepackage{subcaption}
\usepackage{booktabs}
\usepackage{helvet}
\usepackage[normalem]{ulem} 
\usepackage{balance}

\AtBeginDocument{%
  }

\copyrightyear{2026}
\acmYear{2026}
\setcopyright{cc}
\setcctype{by}
\acmConference[KDD '26]{Proceedings of the 32nd ACM SIGKDD Conference on Knowledge Discovery and Data Mining V.2}{August 09--13, 2026}{Jeju Island, Republic of Korea}
\acmBooktitle{Proceedings of the 32nd ACM SIGKDD Conference on Knowledge Discovery and Data Mining V.2 (KDD '26), August 09--13, 2026, Jeju Island, Republic of Korea}
\acmDOI{10.1145/3770855.3819052}
\acmISBN{979-8-4007-2259-2/2026/08}
\newcommand{\Dmax}{$D_{max}$~}
\newcommand{\DC}{$DC_{50}$~}
\newcommand{\pDC}{$pDC_{50}$~}
\newcommand{\ie}{\textit{i.e.}}
\newcommand{\eg}{\textit{e.g.}}

\begin{document}

\title[TACK: A Novel TArgeting Chimeras Knowledge Dataset]{TACK: A Statistical Evaluation of Degradation Activity on a Novel TArgeting Chimeras Knowledge Dataset}

\author{Stefano Ribes}
\email{{ribes}@chalmers.se}
\orcid{0009-0009-2774-8792}
\affiliation{Department of Computer Science and Engineering
  \institution{Chalmers University of Technology and University of Gothenburg}
  \city{Gothenburg}
  \country{Sweden}
}
\author{Nils Dunlop}
\authornote{The first two authors contributed equally to this research.}
\email{{nilsdu}@chalmers.se}
\orcid{0009-0005-1133-8626}
\affiliation{Department of Computer Science and Engineering
  \institution{Chalmers University of Technology and University of Gothenburg}
  \city{Gothenburg}
  \country{Sweden}
}
\author{Rocío Mercado}
\email{{rocio.mercado}@chalmers.se}
\orcid{0000-0002-6170-6088}
\affiliation{Department of Computer Science and Engineering
  \institution{Chalmers University of Technology and University of Gothenburg}
  \city{Gothenburg}
  \country{Sweden}
}

\renewcommand{\shortauthors}{Ribes et al.}

\begin{abstract}
Proteolysis-targeting chimeras (PROTACs) represent a promising therapeutic modality that induces targeted protein degradation by hijacking the ubiquitin-proteasome system.
However, rational PROTAC design remains challenging due to the complex interplay between molecular structure, target proteins, E3 ligases, and the cellular context.
We present TACK, a statistical evaluation of degradation activity on a novel TArgeting Chimeras Knowledge dataset of 3,514 PROTACs and 6,561 degradation endpoints aggregated from three major repositories with standardized molecular representations, protein annotations, and experimental conditions.
Using scaffold-based 5$\times$5 cross-validation, we perform a rigorous statistical comparison of three machine learning methods to predict PROTAC degradation activity across three tasks: \DC and \Dmax regression, and binary activity classification.
Feature ablation demonstrates that cellular context features and simple protein representations rival complex ESM protein embeddings, highlighting the importance of feature engineering over architectural sophistication.
Models trained on the best performing features show that potency ($pDC_{50}$, $R^2=0.66$) is substantially more predictable than maximum degradation ($D_{max}$, $R^2=0.36$).
In activity prediction, statistical tests support that classical methods (XGBoost and MLP) significantly outperform PROTAC-STAN, a domain-specific graph neural network model (ROC-AUC: 0.85 vs. 0.75, $p<0.001$).
Finally, we propose an ensemble-based uncertainty quantification approach showing that prediction variance correlates with prediction error ($pDC_{50}$: Spearman $\rho = 0.36$, $p<0.001$; $D_{max}$: $\rho=0.69$, $p<0.001$), enabling confidence-aware experimental prioritization.
Our findings challenge assumptions about specialized architectures for degradation prediction and provide evidence-based guidance for ML-driven PROTAC assessment.
\end{abstract}



\maketitle

\begin{figure*}[t!]
    \centering
    \includegraphics[width=0.99\textwidth]{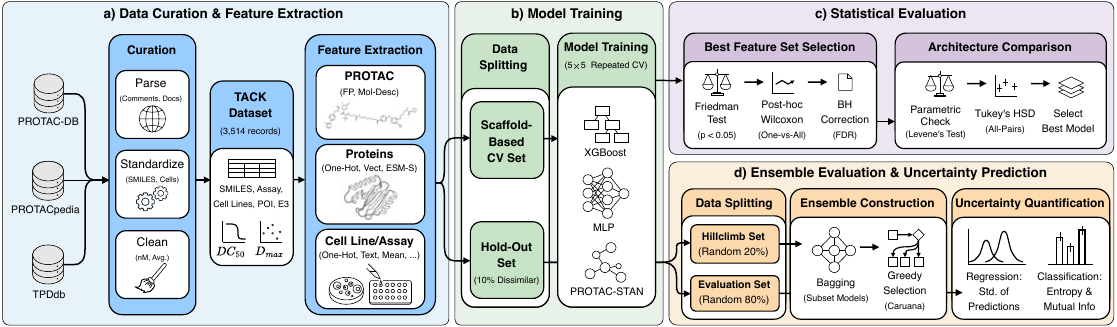}
    \caption{Pipeline summarizing our methodology. (a) Data curation includes merging and cleaning open-source databases to obtain TACK. (b) The TACK dataset is partitioned into a hold-out set, while the rest is used to train models in repeated cross-validation (CV). (c) The models are statistically compared to determine the best feature set and architecture. (d) Ensembles of CV models are used to improve performance and provide epistemic uncertainty.}
    \label{fig:tack_pipeline}
\end{figure*}

\section{Introduction}

Proteolysis-targeting chimeras (PROTACs) represent a paradigm shift in drug discovery, offering a catalytic mechanism to eliminate disease-relevant proteins from cells \cite{sakamoto2001protacs, bekes2022protac}. By recruiting an E3 ubiquitin ligase to a protein of interest (POI), PROTACs harness the cell's native degradation machinery to achieve therapeutic effects at lower doses and against targets long considered undruggable.
Nevertheless, rational design of effective degraders remains challenging due to the complex, non-additive relationship between PROTAC components (warhead, linker, E3 ligand) and degradation efficacy.

Machine learning (ML) offers a compelling path forward for accelerating PROTAC property prediction\cite{gharbi2024comprehensive}, yet three critical gaps hinder progress: \textit{data scarcity}---public datasets contain only a few thousand annotated PROTACs, with >80\% of database entries lacking key activity metrics ($DC_{50}$, $D_{max}$); \textit{lack of standardized benchmarks}---heterogeneous activity thresholds, data splits, and evaluation protocols make fair model comparison difficult; \textit{limited generalizability}---many state-of-the-art models developed for small molecules do not often generalize well to the PROTAC space. Existing databases, including PROTAC-DB \cite{ge2025protac}, PROTACpedia \cite{protacpedia}, PROTAC-Patent-DB \cite{cai2025protac}, and TPDdb \cite{qin2026tpddb}, provide valuable structural and activity data but suffer from inconsistent annotations and coverage, directly impacting ML model training and generalization.

Degradation activity prediction can be framed as either binary classification or regression. Classification enables compound filtering in virtual screening, whereas regression supports compound ranking during lead optimization. Practical tools should support both paradigms. Here we introduce TACK, a curated, ML-ready benchmark for PROTAC degradation activity prediction, addressing the three aforementioned gaps:

\begin{itemize}
    \item The \textit{TACK dataset} harmonizes data from multiple sources with consistent activity thresholds, SMILES standardization, and rigorous quality filters.
    \item The \textit{TACK benchmark} provides a rigorous statistical evaluation of state-of-the-art models, revealing performance gaps and failure modes.
    \item The \textit{TACK ensemble} model is a fast, open-source predictor supporting both classification and regression.
\end{itemize}

All code and data are publicly available at \href{https://github.com/ribesstefano/TACK/}{\uline{this link}}. This work bridges AI \& pharmaceutical science, demonstrating how careful data curation and rigorous benchmarking can advance ML-driven drug discovery for a new therapeutic modality.

\section{Background}

PROTACs were first established as a proof-of-concept by \citet{sakamoto2001protacs} in 2001 using peptide-based recruiters, followed by the first all-small-molecule PROTAC in 2008 \cite{schneekloth2008targeted}. With the field rapidly expanding into clinical trials \cite{chirnomas2023protein}, there is a need for computational tools to screen the vast combinatorial space of linkers, warheads, and E3 ligands that make up PROTACs. Early computational efforts relied on physics-based modeling and molecular simulation to predict stable ternary complex formation \cite{drummond2019, zaidman2020}. While detailed, these methods are too intensive for large-scale library screening, motivating the shift towards ML-based approaches.

Initial predictive models focused on classic ML approaches. \citet{schneider2020} used random forests and gradient boosting machines with molecular descriptors and docking scores as input features for small ligands targeting ER$\alpha$. Similarly, \citet{nori2022} used LightGBM as a reward function within a generative framework. Feature importance analysis consistently identifies PROTAC molecular weight, topological polar surface area, and hydrogen-bond occupancies as key predictors of degradation potency. Recently, deep learning (DL) approaches have tried to learn degradation drivers directly from raw molecular representations. \citet{ribes2024} introduced a framework using 1D and 2D embeddings without 3D dependencies, achieving a test accuracy of 82.6\% for predicting degradation activity. These methods show that high predictive performance is achievable without computationally expensive docking scores, marking a shift toward scalability.

Despite the success of these models, recent work has reintegrated structural insights to capture ternary complex intricacies. DeepPROTACs \cite{li2022deepprotacs} introduced a graph neural network to encode molecular graphs of ligands and generated protein pockets, alongside recurrent neural networks for linker SMILES encoding. However, the model relies on protein-ligand structures, limiting its applicability when high-quality 3D data is sparse. Recently, PROTAC-STAN \cite{chen2025} combined a hierarchical PROTAC encoder with structure-informed POI \& E3 embeddings derived from the protein language model ESM-S \cite{zhang2024esms}. Using a ternary attention network, it explicitly fuses POI–PROTAC–E3 representations, achieving 88.4\% accuracy on its refined dataset, outperforming DeepPROTACs \& \citet{ribes2024}.

Despite these architectural advances, current approaches typically frame PROTAC activity prediction as a binary classification task, simplifying experimental values into `active' or `inactive' labels based on thresholds (commonly \DC<100 nM or \Dmax>80\%) \cite{li2022deepprotacs, ribes2024, chen2025}. This thus fails to distinguish highly potent degraders from marginally active ones. Further, prior work has often relied on random data splitting, which risks overestimating performance on novel chemical structures due to scaffold similarity between train and test sets.
Alongside methodological advances, the scale and quality of available data have evolved from scattered literature reports to centralized repositories. PROTAC-DB \cite{weng2023protacdb} was among the first resources, aggregating experimentally validated degraders from academic literature. Subsequently, the community-driven PROTACpedia \cite{protacpedia} curated >1K degraders. Most recently, TPDdb \cite{qin2026tpddb} expanded the chemical space by mining patent disclosures, cataloging >22K PROTACs alongside emerging modalities like molecular glues. Together, these resources provide the data infrastructure to support the next generation of predictive models.

\section{Methods}

\subsection{TACK Dataset Creation}

\subsubsection{Dataset Curation}

To construct a comprehensive dataset for ML-based PROTAC degradation activity analysis (Figure~\ref{fig:tack_pipeline}a), we aggregated data from three open-access repositories: 6,110 experimentally validated compounds from PROTAC-DB 3.0 \cite{ge2025protac}, 1,189 PROTACs from PROTACpedia \cite{protacpedia}, and a collection of 21,429 PROTACs from TPDdb \cite{qin2026tpddb}. While the initial raw dataset exceeded 29K entries, our subsequent rigorous curation procedure, including filtering for valid \DC \& \Dmax values and chemical standardization, resulted in a high-quality dataset of 3,514 PROTACs.

\begin{table}[t!]
\centering
\caption{Overview of the curated TACK dataset vs. raw data organized by source (\href{https://tpddb.idrblab.net/}{TPDdb}, \href{http://cadd.zju.edu.cn/protacdb/}{PROTAC-DB}, \href{https://protacpedia.weizmann.ac.il/}{PROTACpedia}).}
\label{tab:dataset_comparison}
\begin{center}
\begin{small}
\begin{sc}
\resizebox{\linewidth}{!}{%
\begin{tabular}{lrrrr}
\toprule
\multicolumn{5}{c}{\textbf{Data Overview}} \\
\midrule
\textbf{Statistic} & \textbf{TPDdb} & \textbf{PROTAC-DB} & \textbf{PROTACpedia} & \textbf{TACK} \\
\midrule
Total Records & 22,183 & 9,380 & 1,203 & 6,561 \\
Unique PROTACs & 21,429 & 6,110 & 1,189 & 3,514 \\
Degradation Endpoints & 6,518 & 2,170 & 580 & 6,561 \\
\midrule
POI Targets & 184 & 441 & 79 & 164 \\
E3 Ligases & 8 & 21 & 8 & 9 \\
Cell Lines & 142 & — & 139 & 155 \\
TACK Hold-out Set & 774 & 92 & 53 & 913 \\
\bottomrule
\end{tabular}%
}%
\end{sc}
\end{small}
\end{center}
\vskip -0.1 in
\end{table}

\begin{figure*}[htbp]
    \centering
    \includegraphics[width=0.5\textwidth]{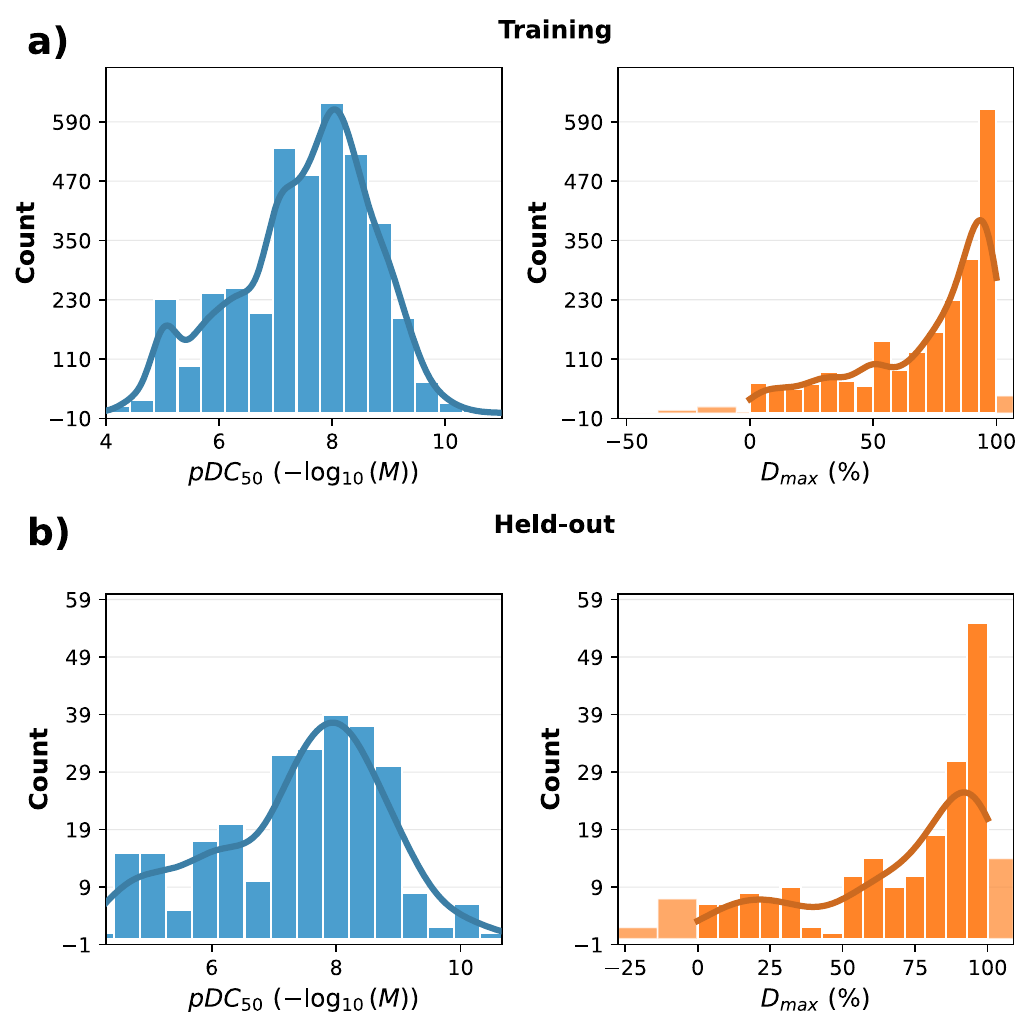}%
    \hspace{-0.01pt}%
   \includegraphics[width=0.5\textwidth]{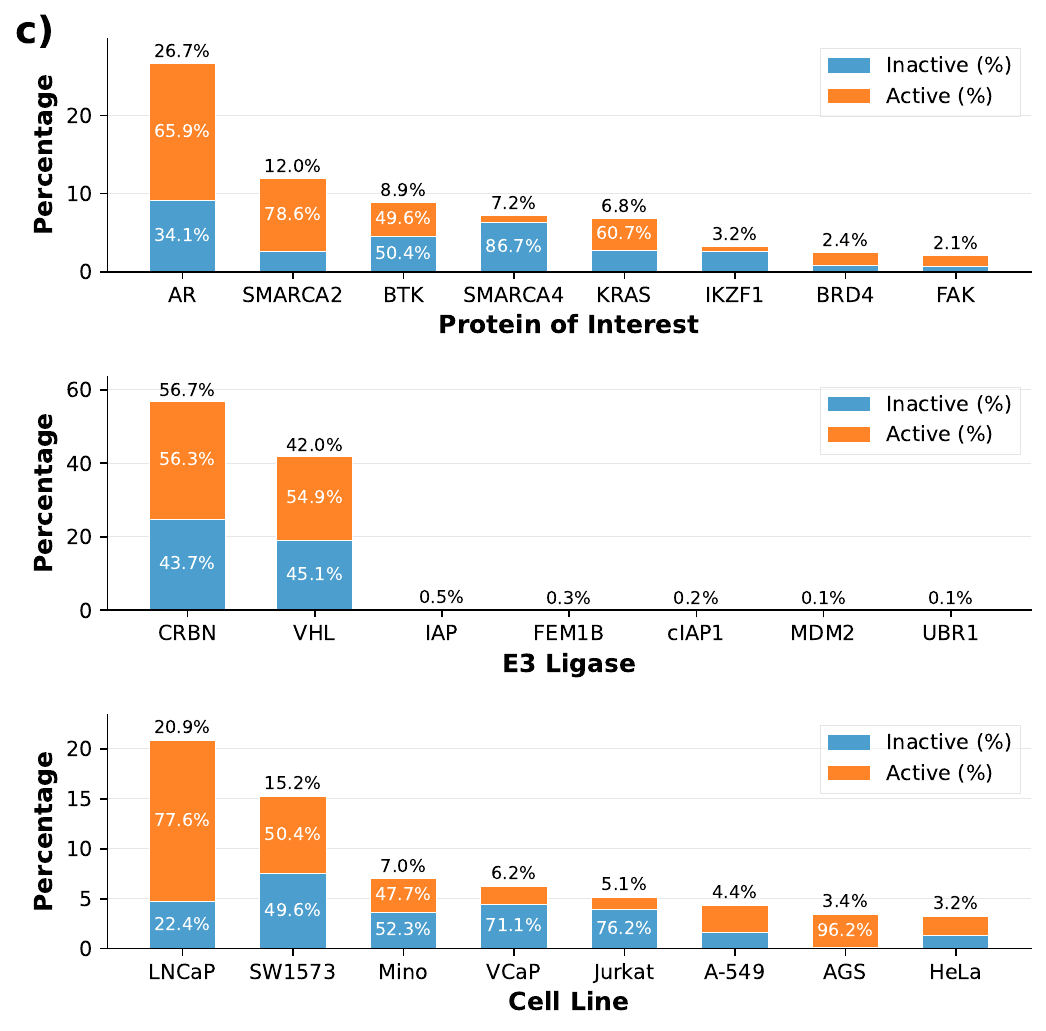}%
    \caption[TACK distributions.]{(a, b) Histograms of standardized degradation activity for the training and hold-out sets, showing the distributions of potency ($pDC_{50}$) and maximal degradation efficacy ($D_{max}$). Opaque bins for \Dmax were clipped between 0 and 100\% for training and evaluation. (c) Stacked bar charts illustrating biological diversity, depicting the top 8 most represented POIs, E3 ligases, and cell lines with the proportions of active versus inactive measurements.}
    \label{fig:tack_distribution}
\end{figure*}

\paragraph{Molecular Standardization.} We implemented a robust data cleaning pipeline designed to standardize molecules and ensure dataset consistency. SMILES representations were canonicalized using RDKit~\cite{rdkit}. POI and E3 ligase names were standardized with their corresponding UniProt identifiers; amino acid sequences for both sets of proteins were retrieved from UniProt where available~\cite{uniprot2025}.

\paragraph{Endpoint Standardization.} \DC \& $D_{max}$ endpoints required special attention due to heterogeneous reporting formats. Numeric values were extracted directly with appropriate unit conversion. All concentrations ($DC_{50}$) were converted to nanomolar units, and \pDC to $DC_{50}$. Values with comparison operators (\eg, ``$>$100 nM'') were treated as their numeric component. However, the operators ($<, >, \leq, \geq$) were stored separately to enable filtering during train-test splitting, allowing entries with inequalities to be excluded from evaluation sets. Range values (\eg, ``10-100 nM'') were converted to their arithmetic mean while storing the original bounds. Categorical grades (A, B, C, D) for activity, common in patent disclosures, were excluded as they required patent-specific mapping to obtain quantitative values.

\paragraph{Assay Standardization.} Cell line names were validated against the Cellosaurus database~\cite{bairoch2018cellosaurus} to obtain standardized identifiers. In TPDdb, the online repository contains no metadata (\eg, cell line, amino acid sequences, assay descriptions) in the downloadable datasets, so we developed a parsing script to extract additional data from the web pages. 
Assay descriptions were standardized (\eg, ``WB'' to ``Western Blot'') to facilitate grouping by experimental method. For each \Dmax measurement, the treatment concentration was extracted from multiple sources including explicit metadata fields and parsed assay descriptions, where available. Patent information was sourced from the original patent database tables. For PROTAC-DB, the primary challenge was parsing assay descriptions that encoded multiple experimental conditions within single text entries. Descriptions such as ``\textit{Degradation of BRD4 short/long in HeLa cells after 24 h treatment}'' required systematic extraction of target proteins, cell lines, and treatment times. We developed parsing functions to handle slash-separated multi-target entries, protein mutation annotations, and various time formats. Value columns containing multiple measurements (\eg, ``\textit{0.081/0.14/0.53}'') were split into individual entries. PROTACpedia required parsing of free-text comment fields where experimental values were embedded within narrative descriptions (\eg, ``\textit{DC50 is 0.86 nM in LNCaP, 0.76 nM in VCaP}''). To extract cell-line-specific measurements, we implemented pattern-matching handlers for comments, which often described multi-protein degradation panels. Entries containing multiple targets within a single row were excluded for consistency.

Table \ref{tab:dataset_comparison} and Figure \ref{fig:tack_distribution} summarize the characteristics of the TACK dataset. The aggregation and filtering processes resulted in 6,561 total experimental endpoints comprising 4,184 potency ($DC_{50}$) and 2,377 efficacy ($D_{max}$) measurements. The dataset covers a broad but consistent activity range in both the training/validation (Figure \ref{fig:tack_distribution}a) and held-out (Figure \ref{fig:tack_distribution}b) sets. We report the distribution of $pDC_{50}$, the negative logarithm (base 10) of \DC in molar concentration, to better visualize potency. \Dmax values are mostly concentrated at high degradation levels, with a peak around 90–100\%, while \pDC values are more normally distributed with a slight skew towards lower potency. These trends indicate a bias towards effective degraders, though PROTAC potency varies widely. The dataset spans a diverse biological landscape totaling 164 distinct POIs and 155 cell lines, with the eight most common POIs, E3 ligases, and cell lines detailed in Figure \ref{fig:tack_distribution}c. Overall, activity labels are fairly balanced, with approximately 55\% of entries classified as active in both the training/validation and held-out sets (Appendix \ref{app:cv_distributions}).

\subsubsection{Scaffold Clustering and Hold-out Set}
\label{subsec:scaffold_clustering}

We constructed a hold-out test set isolating $\sim$10\% of the curated TACK data based on structural dissimilarity.
Using RDKit, we calculated 512-bit Morgan8 fingerprints for all PROTACs in TACK and calculated pairwise Tanimoto distances.
We averaged the distances for each PROTAC to obtain a single dissimilarity score.
The hold-out set was formed by selecting $\sim$10\% of points with the highest dissimilarity score to simulate a realistic scenario in which models must predict degradation activity for novel PROTACs that differ structurally from those seen during training.
This hold-out set was kept separate from the training and validation steps. For the remaining data, we clustered PROTACs using a scaffold-based grouping strategy based on Murcko scaffolds~\cite{bemis1996properties}.
This grouping guarantees that during cross-validation (CV), no PROTACs sharing the same scaffold would appear in both training and validation sets, thereby providing a stringent evaluation of model generalization to unseen chemical structures.
Compared to random splitting, which can lead to optimistic performance estimates, scaffold splitting offers a more realistic assessment of model capabilities in drug development~\cite{ash2025practically}.

\subsection{Prediction Tasks}

\subsubsection{Task Definition}

Our analysis focuses on three tasks related to degradation activity: (1) potency ($pDC_{50}$) regression, (2) efficacy ($D_{max}$) regression, and (3) binary classification of degradation activity. For activity classification, we define a data point as `active' if it reports \Dmax > 80\% \& \DC < 100 nM, following thresholds in \citet{li2022deepprotacs}.
\Dmax \& \DC values are independently scaled with a quantile transformation~\cite{gilchrist2000statistical} prior to regression.
Since such normalization is data-dependent, we fit the transformation only on the training set and apply it to the validation and hold-out sets.

\subsubsection{Feature Extraction}
\label{sec:feature_extraction}

PROTAC ternary complexes consist of the POI, the E3 ligase, and the PROTAC molecule itself; we describe each of these three components by its own feature. For POIs and E3 ligases, we extract features based on their amino acid sequences: either as 1- and 2-grams with normalized counts (referred to as ``\textsc{Vec}''), or as ESM-S precomputed embeddings~\cite{zhang2024esms}. We compress ESM-S embeddings via PCA to retain 90\% of their variance, bringing the original 1280 embeddings dimension down to 44 and 7 for POI- and E3 ligase-related embeddings, respectively; we refer to these compressed embeddings simply as ``\textsc{ESM-S}''.
For PROTACs, we compute either 512-bit Morgan8 fingerprints (referred to as ``\textsc{FP}'') or RDKit descriptor fingerprints (217 descriptors; referred to as ``\textsc{Desc}'') using RDKit.
In some models, we use both Morgan8 and descriptor fingerprints, which we refer to as the ``\textsc{Desc+FP}'' feature set.
Additionally, experimental conditions such as cell line, assay type, and treatment time influence degradation activity.
For cell lines, we extract a textual description of the cell line and embed it using a pre-trained sentence transformer~\citep[Appendix B]{ribes2024} (referred to as ``\textsc{Text}'').
The assay type (\eg, Western blot, ELISA, HiBit), is encoded with one-hot encoding.
The treatment time is instead encoded as a single numerical feature representing the duration of the assay treatment in hours. For assay conditions, we handle missing values as follows: embed the string ``\textit{Unknown cell line.}'' for missing cell lines, impute missing treatment times with the mean of the training set, and use a zero vector for missing assay types.
Finally, we also experimented with: a \textit{simple} one-hot encodings of the POI gene name, the E3 ligase name, and/or the cell line (referred to as ``\textsc{OneHot}'' in all cases), and a representation that mimics the input features of PROTAC-STAN, \ie, molecular fingerprints and ESM-S embeddings without applying PCA for POI and E3 ligase.
The complete list of features is reported in Appendix \ref{app:feature_sets}, Table \ref{tab:feature_sets}.

\subsection{Model}

\subsubsection{Selected Models}

In this work, we evaluate three model architectures: XGBoost, a multi-layer perceptron (MLP), and PROTAC-STAN~\cite{chen2025}, the latter only used for the binary classification task as a literature baseline. Different combinations of feature sets (Section \ref{sec:feature_extraction}) are evaluated with both the XGBoost and MLP models. Following concatenation of the selected input features, all models predict a single scalar value, corresponding to either a regression or classification target. See Appendix \ref{app:model_details} for model details.

\subsubsection{CV}

We perform a repeated group 5$\times$5 CV (later referred to as \textit{repeated CV}) on the TACK dataset, excluding the hold-out set, for a rigorous statistical evaluation as described in~\citet{ash2025practically}.
The inner CV loop splits the data into 5 folds, where each fold is set aside once for validation while the remaining 4 folds are used for training.
The outer CV loop repeats this process 5 times with different random initialization seeds, leading to a total of 25 unique folds and 25 models trained per model class and task.
As detailed in Section \ref{subsec:scaffold_clustering}, groups are defined by the Murcko scaffolds of the PROTAC molecules.
Figure \ref{fig:cv_distributions}, Appendix \ref{app:cv_distributions}, shows the distribution of the label values in the different folds.
See Appendix \ref{app:hyperparameters} for a detailed description of the hyperparameter optimization procedure using CV.

\subsection{Statistical Evaluation}
\label{sec:statistical_evaluation}

To compare model configurations and identify optimal architectures for each prediction task, we used a hierarchical statistical framework designed to control false discovery rates while maintaining statistical power across multiple comparisons.
For regression tasks, we selected root mean squared error (RMSE) as our primary evaluation metric, while for the binary classification task, we use the area under the receiver operating characteristic curve (ROC-AUC). All metrics are computed on the 5$\times$5 CV scheme, yielding 25 independent performance estimates per configuration.

\subsubsection{Best Feature Set}
\label{sec:identify_best_features}

A key challenge in applying ML to activity prediction is determining which combination of model and molecular/assay representations yields the best performance. We evaluated 10 feature combinations (Appendix \ref{app:feature_sets}) for both XGBoost and MLP architectures across each of the three prediction tasks.

To identify top-performing configurations while accounting for variance across CV folds, we employed a rigorous two-stage statistical testing procedure. First, we applied the Friedman test~\cite{friedman1937} to assess whether significant performance differences exist among configurations. Upon detecting significant omnibus differences ($p<0.05$), we conducted post-hoc pairwise comparisons using Wilcoxon signed-rank tests~\cite{wilcoxontests}: we identified the configuration with the best mean performance in the validation set as the control method and compared all other configurations against this top-performing control.
We controlled the false discovery rate (FDR) using Benjamini-Hochberg (BH) correction~\cite{benjamini1995controlling} with $\alpha = 0.05$, chosen over family-wise error rate methods (\eg, Bonferroni) for its superior statistical power in high-dimensional comparison settings~\cite{ash2025practically}. Configurations not rejected by the BH-corrected tests were deemed statistically equivalent to the best method. This principled approach ensures that our feature recommendations generalize beyond random fold variations, an important consideration for practitioners applying these models to novel PROTAC designs.

\subsubsection{Model Comparison}
\label{sec:model_comparison}

Having identified optimal feature configurations, we compared the XGBoost, MLP, and PROTAC-STAN models on the validation folds. For classification, when multiple feature sets were statistically equivalent, we selected the simplest configuration (\eg, one-hot encodings over sequence embeddings) to prioritize computational efficiency and limit the number of comparisons, thereby avoiding inflated Type I error rates (\eg, p-hacking).

The comparison procedure mirrored that used for feature selection, but in model comparison we only compare three architectures instead of 10 feature combinations, thus favoring the use of other statistical tests.
We verified homogeneity of variances using Levene's test~\cite{levene1960robust} and assessed normality visually, as the 5$\times$5 CV metrics should normally distribute due to the central limit theorem (Appendix \ref{app:normality_diagnostic}). When assumptions were satisfied, we applied Tukey's HSD test~\cite{tukey1949comparing}, which provides tighter confidence intervals than non-parametric alternatives in small-sample comparisons.

\subsection{Ensembling and Uncertainty Quantification}

Reliable uncertainty estimates are critical for deploying ML models in scientific discovery workflows, where predictions guide expensive experimental validation. We construct an ensemble of the 5$\times$5 CV models and quantify epistemic uncertainty---the uncertainty arising from limited training data and model specification---to help practitioners identify predictions requiring additional validation.

\subsubsection{Ensemble Selection}
\label{sec:caruana}

We employed Caruana's greedy forward selection algorithm~\cite{caruana2004}, which iteratively builds an ensemble by adding models that maximize performance on a dedicated selection set. To ensure unbiased evaluation, we partitioned the held-out set into a selection subset (19.9\%) for ensemble construction and an evaluation subset (80.1\%) for final assessment. For each task (\pDC regression, \Dmax regression, and binary activity classification), we considered all 500 candidate models (10 feature configurations, both XGBoost and MLP models, $\times$ 25 CV folds), including lower-performing models to promote diversity. 

The algorithm initializes with the top 5 models and iteratively adds the candidate that most improves selection set performance (RMSE for regression, log loss for classification). Models may be selected multiple times, with final weights normalized to sum to 1. We applied bagging (10 random samples of 50\% of available models) to reduce sensitivity to selection set variability.

We evaluated two pools of candidate models for ensemble selection: (1) \textit{model-level} selection, where candidates include all 500 individual CV fold models; and (2) \textit{architecture-level} selection, where candidates are the 20 unique feature configurations with predictions pre-averaged across their respective CV folds. We compared against three baselines: the single best model, uniform averaging of all 500 models, and uniform averaging of the 25 CV folds for the best model (as defined by model type and feature set).

\subsubsection{Uncertainty Quantification}

We derive uncertainty estimates from inter-model disagreement within the ensemble. For regression, we calculated the standard deviation ($\sigma$) across member predictions.
For classification, we calculate predictive entropy and decompose it to isolate mutual information as a measure of epistemic (model) uncertainty. To assess calibration, we measure Spearman correlation between uncertainty and absolute prediction error for regression tasks; well-calibrated models should exhibit positive correlation, indicating that uncertain predictions correspond to larger errors. For classification, we report expected and maximum calibration error (ECE/MCE)~\cite{pavlovic2025understandingmodelcalibration}, quantifying the alignment between predicted probabilities and observed frequencies. These calibration metrics are essential for accurate compound prioritization using our models: miscalibrated confidence could lead to suboptimal resource allocation in PROTAC design campaigns.

\section{Results}

\begin{table}[t!]
\centering
\caption{Best feature set configurations for XGBoost and MLP models. We report the statistically equivalent feature combinations per task and per model. The metrics are calculated as the mean of the scores computed on the validation folds. Shorthand for feature sets is defined in Sec. \ref{sec:feature_extraction}.}
\label{tab:best_features_configurations}
\begin{center}
\begin{small}
\begin{sc}
\resizebox{\linewidth}{!}{%
\begin{tabular}{llccccr}
\toprule
\textbf{Task} & \textbf{Model} & \multicolumn{4}{c}{\textbf{Best Feature Set Configuration}} & \multirow{2}{*}{\shortstack[l]{\textbf{Best Val.}\\\textbf{Mean}}} \\
\cmidrule(lr){3-6}
& & \textbf{Cell} & \textbf{E3} & \textbf{POI} & \textbf{Mol} & \\
\midrule
\multirow{4}{*}{\shortstack[l]{\pDC\\(RMSE)}} 
& \small{MLP}$^{(\dag)}$ & \textbf{OneHot} & \textbf{OneHot} & \textbf{OneHot} & \textbf{Desc} & \textbf{0.770} $(\downarrow)$ \\
& \small{MLP} & OneHot & OneHot & Vec & Desc & 0.787 $(\downarrow)$ \\
& \small{XGB} & Text & ESM-S-PCA & ESM-S-PCA & Desc & 0.781 $(\downarrow)$ \\
& \small{XGB} & Text & OneHot & Vec & Desc+FP & 0.785 $(\downarrow)$ \\
\midrule
\multirow{6}{*}{\shortstack[l]{\Dmax\\(RMSE)}} 
& \small{MLP} & OneHot & OneHot & OneHot & Desc & 20.589 $(\downarrow)$ \\
& \small{MLP} & OneHot & OneHot & Vec & Desc & 20.689 $(\downarrow)$ \\
& \small{XGB} & Text & ESM-S-PCA & ESM-S-PCA & Desc+FP & 18.452 $(\downarrow)$ \\
& \small{XGB} & Text & ESM-S-PCA & ESM-S-PCA & Desc & 18.546 $(\downarrow)$ \\
& \small{XGB} & Text & OneHot & OneHot & Desc & 18.514 $(\downarrow)$ \\
& \small{XGB}$^{(\dag)}$ & \textbf{Text} & \textbf{OneHot} & \textbf{Vec} & \textbf{Desc} & \textbf{18.340} $(\downarrow)$ \\
\midrule
\multirow{6}{*}{\shortstack[l]{Activity\\(ROC-AUC)}} 
& \small{MLP}$^{(\ddag)}$ & OneHot & OneHot & OneHot & Desc & 0.805 $(\uparrow)$ \\
& \small{XGB}$^{(\ddag)}$ & \textbf{Text} & \textbf{ESM-S-PCA} & \textbf{ESM-S-PCA} & \textbf{Desc+FP} & \textbf{0.851} $(\uparrow)$ \\
& \small{XGB} & Text & ESM-S-PCA & ESM-S-PCA & Desc & 0.850 $(\uparrow)$ \\
& \small{XGB} & Text & OneHot & Vec & Desc & 0.848 $(\uparrow)$ \\
& \small{XGB} & Text & OneHot & Vec & Desc+FP & 0.847 $(\uparrow)$ \\
& \small{XGB} & Text & OneHot & OneHot & Desc & 0.847 $(\uparrow)$ \\
\bottomrule
\\
\multicolumn{7}{l}{\normalfont ${(\dag)}$: Models used to plot hold-out set predictions in Fig. \ref{fig:results}a.} \\
\multicolumn{7}{l}{\normalfont ${(\ddag)}$: Models selected for architecture comparison with PROTAC-STAN in Fig. \ref{fig:results}b.}
\end{tabular}
}%
\end{sc}
\end{small}
\end{center}
\vskip -0.2 in
\end{table}

\begin{figure*}[t!]
    \centering
    \includegraphics[width=\textwidth]{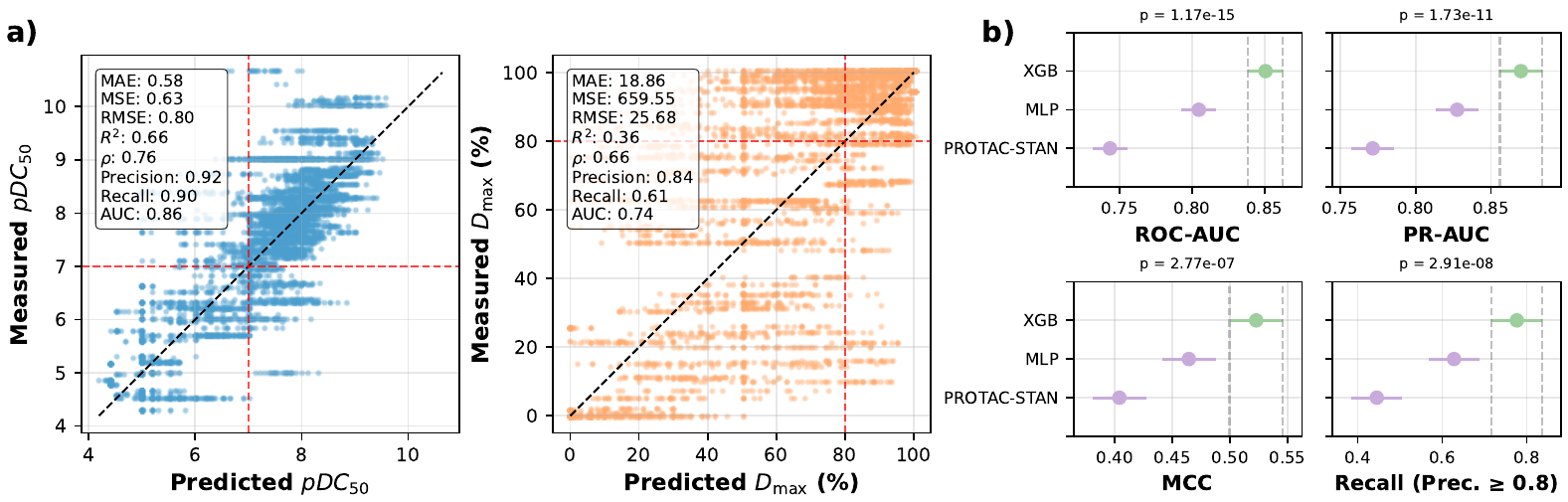}
    \caption{(a) Parity plots using MLP$^{(\dag)}$ and XGBoost$^{(\dag)}$ for \pDC and $D_{max}$, respectively, on the hold-out test set (25 CV fold models). (b) Performance comparison for XGBoost$^{(\ddag)}$, MLP$^{(\ddag)}$, PROTAC-STAN on the binary activity classification task across four metrics. Points show means with 95\% confidence intervals from 25 CV folds. Non-overlapping intervals indicate significant differences (Tukey HSD, $p < 0.05$). Omnibus ANOVA p-values shown above each panel. Precise effect size values are reported in Figure \ref{fig:effect_size_bin_val}.}
    \label{fig:results}
\end{figure*}

\subsection{TACK Dataset}

The TACK dataset is summarized in Table \ref{tab:dataset_comparison} \& Figure \ref{fig:tack_distribution}. Among the 164 POI targets, the androgen receptor (AR, 26.7\%), SMARCA2 (12.0\%), and BTK (8.9\%) are the most frequently studied, with the top five POIs accounting for 61.6\% of all endpoints, reflecting the field's focus on oncology targets. E3 ligase diversity remains limited, with CRBN and VHL comprising 98.7\% of all measurements. Cell line selection is more heterogeneous, with LNCaP (20.9\%), SW1573 (15.2\%), and Mino (7.0\%) being the most common among 155 unique lines. Overall, 23.8\% of PROTACs have multiple endpoints and 55.1\% meet activity thresholds (\DC $\leq$ 100 nM and \Dmax $\geq$ 80\%).

\subsection{Statistical Evaluation Results}

\subsubsection{Best Feature Set}

We evaluated feature configurations across XGBoost and MLP architectures using 5$\times$5 CV on TACK, applying the statistical framework from Section~\ref{sec:identify_best_features}. Table~\ref{tab:best_features_configurations} summarizes configurations identified as statistically equivalent to the best-performing setup after Benjamini-Hochberg correction.

\textbf{XGBoost}. Friedman tests detected highly significant differences across feature configurations for \pDC ($\chi^2=174.77$, $p=6.2\times10^{-33}$), \Dmax ($\chi^2=131.41$, $p=6.1\times10^{-24}$), and binary classification ($\chi^2=122.2625$, $p=4.6\times10^{-22}$). The optimal XGBoost configuration for \pDC combined cell line \textsc{Text} embeddings, \textsc{ESM-S-PCA} encodings for both E3 ligase and POI, and \textsc{Desc} for PROTACs (val. RMSE $=0.781$). For \Dmax prediction, multiple XGBoost feature sets achieved statistical equivalence after Benjamini-Hochberg correction, with the best employing cell \textsc{Text}, E3 \textsc{OneHot}, POI \textsc{Vec}, and \textsc{Desc} (RMSE $=18.34$).
On binary classification, BH-corrected tests yielded four statistically equivalent XGBoost configurations (val. ROC-AUC $=0.847-0.851$), with the best using \textsc{ESM-S-PCA} protein embeddings (cell: \textsc{Text}, E3 and POI: \textsc{ESM-S-PCA}, mol: \textsc{Desc+FP}; ROC-AUC $=0.851$).

\textbf{MLP}. MLPs showed distinct patterns across tasks. For \pDC, the Friedman test indicated significant configuration differences ($\chi^2=129.77$, $p=1.3\times10^{-23}$), with two statistically equivalent feature sets emerging, both using simple \textsc{OneHot} encodings. Notably, the best MLP configuration (cell: \textsc{OneHot}, E3: \textsc{OneHot}, POI: \textsc{OneHot}, mol: \textsc{Desc}; RMSE $=0.770$) slightly outperformed the best XGBoost setup by 1.4\%. For \Dmax, two MLP configurations survived post-hoc testing ($\chi^2=105.25$, $p=1.4\times10^{-18}$), with the best achieving RMSE $=20.59$---exhibiting 12.3\% higher error than the best XGBoost alternative.
For binary classification, Friedman tests ($\chi^2=102.95$, $p=4.0\times10^{-18}$) identified a unique best MLP configuration (cell: \textsc{OneHot}, E3: \textsc{OneHot}, POI: \textsc{OneHot}, mol: \textsc{Desc}; ROC-AUC $=0.805$), trailing the top XGBoost model by 5.4\%.

\subsubsection{Generalization to Hold-out Set}

To assess generalization performance, we evaluated the best performing MLP and XGBoost configurations (Table~\ref{tab:best_features_configurations}, marked with $\dag$) on the hold-out test set, aggregating predictions from all 25 CV fold models (Figure~\ref{fig:results}a). For prediction of $pDC_{50}$, the MLP model achieved on average strong performance (test MAE $=0.58$, MSE $=0.63$, $R^2=0.66$, Spearman $\rho=0.76$), with the parity plot showing well-calibrated predictions throughout the entire \pDC concentration range and minimal systematic bias. When using a binary activity threshold of 100 nM, as reported in DeepPROTACs~\cite{li2022deepprotacs}, the model exhibited high sensitivity in identifying active compounds (recall $=0.90$) while maintaining high specificity (precision $=0.92$, ROC-AUC $=0.86$), indicating effective separation of potent degraders from weak or inactive compounds. For \Dmax predictions, XGBoost showed substantially weaker performance (MAE $=18.86$, MSE $=659.55$, $R^2 = 0.36$, Spearman $\rho = 0.66$), with the parity plot revealing considerable scatter and only a modest rank-order correlation. Despite the low coefficient of determination, the model retained a moderate discriminative ability to classify high vs. low degradation at a threshold of 80\%~\cite{li2022deepprotacs} (ROC-AUC $=0.74$, precision $=0.84$, recall $=0.61$), suggesting that while precise quantification of \Dmax remains challenging, coarse categorical predictions (\eg, strong degraders vs. weak degraders) are feasible.

\subsubsection{Model Comparison for Activity Classification}

Figure~\ref{fig:results}b shows statistical comparison of the three models on the folds validation sets using the best-performing configurations (Table~\ref{tab:best_features_configurations}, marked with $\ddag$).
ANOVA detected significant architecture effects for all metrics: ROC-AUC ($p = 1.17 \times 10^{-15}$), PR-AUC ($p = 1.73 \times 10^{-11}$), MCC ($p = 2.77 \times 10^{-7}$), and recall at 80\% precision ($p = 2.91 \times 10^{-8}$).
Tukey's HSD post-hoc tests with family-wise error rate correction identified XGBoost as the best-performing model across all four metrics (ROC-AUC: 0.851, PR-AUC: 0.870, MCC: 0.523, Recall@80\%P: 0.777), significantly outperforming both MLP and PROTAC-STAN on each.
MLP in turn significantly outperformed PROTAC-STAN on all four metrics, with the largest margin on recall at 80\% precision ($\Delta = 0.183$, $p_{\text{adj}} = 0.001$).

\subsection{Ensemble Results}

Table~\ref{tab:ensemble_performance_uncertainty} shows the performance of the ensemble methods and baselines on the isolated held-out set.
On $pDC_{50}$, the best single model achieved the lowest test RMSE (0.633), with the Caruana ensemble (RMSE$=0.672$, 33 models) ranking second and outperforming both the uniform average (0.683) and best model ensemble (0.685). On $D_{max}$, the best single model again achieved the lowest RMSE (20.94), with the Caruana ensemble (RMSE$=21.32$, 22 models) ranking second and outperforming the best model ensemble (22.36).
For binary classification, the best single model achieved the lowest log loss (0.295), and ensemble strategies performed worse: best model ensemble (0.312, 25 models), Caruana ensemble (0.317, 18 models), and uniform average (0.380, 500 models).

\subsubsection{Uncertainty Quantification}
\label{sec:uq}

Ensemble uncertainty estimates demonstrate meaningful correlations with prediction errors across regression tasks (Table~\ref{tab:ensemble_performance_uncertainty}). For \pDC prediction, the uniform average of all models achieved the strongest uncertainty-error correlation (Spearman $\rho = 0.355$, $p < 0.001$), while the Caruana ensemble showed a weaker but significant correlation ($\rho = 0.207$, $p < 0.01$). The Caruana ensemble's prediction standard deviation averaged 0.361 \pDC units, with 46.5\% of the test samples falling within $\pm$1$\sigma$ intervals and 88.0\% within $\pm$3$\sigma$. These coverage values fall below the theoretical expectations of 68.3\%, 95.5\%, and 99.7\% for normally distributed errors, suggesting systematic underestimation of uncertainty by approximately 20-30\%. For $D_{max}$, the uniform average exhibits the highest correlation ($\rho=0.694$) alongside the best coverage at 2$\sigma$ (80.5\%), while the Caruana ensemble shows more balanced performance ($\rho=0.543$, 47.9\% coverage at 1$\sigma$). The best model ensemble achieves the lowest uncertainty spread ($\bar{\sigma}=7.59$) but severely underestimates coverage (21.3\% at 1$\sigma$), highlighting a tradeoff between uncertainty magnitude and correlation strength.

For binary classification, the ECE ranges from 10.0\% (\textsc{Best Ensem.}) to 12.1\% (\textsc{Caruana}), indicating moderate calibration with average deviations around 10\%. Unlike regression tasks, MCE values are substantially lower (28.2--35.6\%), suggesting that high-confidence predictions are more reliable for binary classification than for regression.

\begin{table}[t!]
\centering
\caption{Ensemble performance and uncertainty quantification on held-out test set. Performance evaluated on 80\% evaluation subset (20\% used for Caruana selection). Uncertainty-error correlation: Spearman $\rho$ between std and absolute error. Coverage: fraction within k$\sigma$ (expected: 68.3\%, 95.5\%, 99.7\%). Except for \textsc{Best Single}, all other methods are ensembles: \textsc{Uniform} averages all 500 models; \textsc{Best Ensem.} averages the 25 best models; \textsc{Caruana} is described in Sec. \ref{sec:caruana}}
\label{tab:ensemble_performance_uncertainty}
\begin{small}
\begin{sc}
\resizebox{\linewidth}{!}{%
\begin{tabular}{llcc|cccccc}
\toprule
\multicolumn{4}{c|}{\textbf{Performance}} & \multicolumn{6}{c}{\textbf{Uncertainty Quantification}} \\
\cmidrule(lr){1-4} \cmidrule(lr){5-10}
\textbf{Task} & \textbf{Method} & \textbf{Metric} & \textbf{N} & $\bar{\sigma}$ & $\rho$ & \textbf{1$\sigma$} & \textbf{2$\sigma$} & \textbf{3$\sigma$} & \textbf{ECE/MCE} \\
\midrule
\multirow{4}{*}{\shortstack[l]{$pDC_{50}$\\(RMSE)}} 
& Best Single & \textbf{0.633} & 1 & -- & -- & -- & -- & -- & -- \\
& Uniform & 0.683 & 500 & 0.365 & \textbf{0.355**} & 42.4 & \textbf{78.3} & \textbf{91.2} & -- \\
& Best Ensem. & 0.685 & 25 & 0.284 & 0.303** & 33.6 & 64.5 & 82.0 & -- \\
& Caruana & 0.672 & 33 & 0.361 & 0.207* & \textbf{46.5} & 70.5 & 88.0 & -- \\
\midrule
\multirow{4}{*}{\shortstack[l]{$D_{max}$\\(RMSE)}} 
& Best Single  & \textbf{20.94} & 1   & --    & -- & -- & -- & -- & -- \\
& Uniform      & 22.92 & 500 & 12.53 & \textbf{0.694**} & 37.3 & \textbf{80.5}  & \textbf{91.7}  & -- \\
& Best Ensem.  & 22.36 & 25  & 7.59  & 0.614** & 21.3 & 49.7 & 74.0 & -- \\
& Caruana      & 21.32  & 22  & 12.47 & 0.543** & \textbf{47.9} & 78.1 & 91.1 & -- \\
\midrule
\multirow{4}{*}{\shortstack[l]{Activity\\(LogLoss)}}
& Best Single  & \textbf{0.295} & 1   & -- & -- & -- & -- & -- & -- \\
& Uniform      & 0.380          & 500 & -- & -- & -- & -- & -- & 11.4 / \textbf{28.2} \\
& Best Ensem.  & 0.312          & 25  & -- & -- & -- & -- & -- & \textbf{10.0} / 33.6 \\
& Caruana      & 0.317          & 18  & -- & -- & -- & -- & -- & 12.1 / 35.6 \\
\bottomrule
\multicolumn{10}{l}{\normalfont \footnotesize **$p<0.001$; *$p<0.01$. Coverage, ECE, and MCE are expressed in \%.}
\end{tabular}
}%
\end{sc}
\end{small}
\vskip -0.2in
\end{table}

\section{Discussion}

Our systematic evaluation of ML approaches for PROTAC degradation reveals several insights relevant to both the AI/ML and drug discovery communities. We discuss here key findings, their implications, and limitations that inform future research directions.

\subsection{Feature Representation \& Simplicity}

A striking finding from our feature selection analysis (Table \ref{tab:best_features_configurations}) is that simple encodings (\textsc{OneHot}, \textsc{Vec}) often matched or approached the performance of computationally expensive ESM-S embeddings, particularly when paired with XGBoost. This suggests that for PROTAC activity prediction, lightweight representations may suffice when paired with tree-based models. Further, the task-specific differences in optimal feature sets warrant additional discussion.
For both \pDC and \Dmax predictions, the best configurations included descriptor-only, whereas \textsc{Desc+FP} representations performed comparably. This divergence may reflect the distinct physicochemical determinants of these endpoints, expanded on in the next subsection.
The success of simple features comes with an important caveat regarding generalization. One-hot encodings, by construction, cannot represent proteins unseen during training: an unknown POI has no meaningful bit mapping. This limits applicability to novel targets, a critical consideration given that expanding the ``druggable'' proteome is a key goal of PROTAC research.

\subsection{Asymmetry in Endpoint Predictability}

The substantial gap in predictive performance between \pDC ($R^2 = 0.66$) and \Dmax ($R^2 = 0.36$) reveals a fundamental asymmetry in PROTAC predictability. Potency ($pDC_{50}$) is considerably easier to learn from molecular and contextual features than maximal degradation efficacy ($D_{max}$). This asymmetry likely reflects the underlying biology. \DC is primarily governed by ternary complex formation kinetics and PROTAC-target binding affinity---properties with clearer structure-activity relationships amenable to ML modeling. In contrast, \Dmax depends on a cascade of factors: E3 ligase expression levels, target protein localization, proteasome availability, and the presence of competing binding partners that may occlude the target from PROTAC engagement \cite{cardno2025cellular}. These factors vary across cell lines and experimental conditions in ways that the explored molecular representations do not capture.
From a practical standpoint, this finding suggests that binary classification (active/inactive) and \pDC regression are more reliable for computational screening, while \Dmax predictions should be interpreted with greater caution.

\subsection{Tree-Based Models Outperform DL}

XGBoost outperformed both MLP and PROTAC-STAN across nearly all tasks, falling short of MLP by 1.4\% on predicting $DC_{50}$, a result that merits examination given the recent emphasis on DL for molecular property prediction. We hypothesize several contributing factors. With <4K unique labeled PROTACs in TACK, the dataset may be insufficient for deep architectures to learn robust representations. Tree-based methods are well-established as strong performers on small-to-medium tabular datasets \cite{jiang2021could}, where their design provides effective regularization. The prevalence of activity thresholds in the literature (\eg, 100 nM for $DC_{50}$, 80\% for $D_{max}$) creates natural decision boundaries that tree-based models can exploit through their hierarchical structure. Neural networks must learn these boundaries implicitly, requiring more data.

PROTAC-STAN, despite its attention-based architecture designed specifically for PROTAC activity prediction, underperformed relative to simpler tree-based models on TACK. This performance gap likely reflects a mismatch in model complexity and the available data size, or suboptimal hyperparameter transfer from the original setting. Notably, PROTAC-STAN was fine-tuned on the authors' curated PROTAC-DB subset, where it achieved 88\% accuracy on held-out data~\cite{chen2025}. The worse performance we observe on TACK---which incorporates structurally diverse compounds from TPDdb beyond the original PROTAC-DB distribution---suggests potential overfitting to the fine-tuning domain.
We did not evaluate DeepPROTACs~\cite{li2022deepprotacs} because it requires pre-computed binding pocket structures for both the POI and E3 ligase, information which is unavailable for the majority of TACK entries. This exclusion highlights a broader challenge in the field: methods incorporating 3D inputs may offer richer representations, but face significant data availability constraints in practice. As structure prediction tools continue to mature \cite{dunlop2025predicting}, revisiting structure-aware approaches on new benchmarks represents a promising direction of future work.

\subsection{Uncertainty Quantification and Data Quality}

Our ensemble-based uncertainty estimates showed positive correlation with prediction error (Sec. \ref{sec:uq}), indicating reasonable calibration. However, the modest magnitude of these correlations suggests that epistemic uncertainty (reducible through additional data or improved models) may not fully explain prediction errors, especially in the case of \pDC prediction.
This observation points to substantial aleatoric uncertainty inherent in the data itself. Sources can include: assay variability across laboratories and protocols; cell line-specific effects not captured by current encodings; and measurement noise in reported endpoint values. If aleatoric uncertainty dominates, even perfect models cannot achieve high predictive accuracy without improved data quality.
The broader dataset context supports this interpretation: of the nearly 30K experimentally characterized PROTACs across public databases, only $\sim$9K ($\sim$30\%) have associated degradation measurements. The remaining compounds lack quantitative activity data, representing both a limitation and an opportunity. Active learning methods that iteratively prioritize informative compounds for experimental characterization could efficiently expand the labeled dataset while maximizing model improvement per experiment \cite{van2024traversing}.

\section{Limitations and Ethical Considerations}

Several limitations should be considered when interpreting our results. \textit{Generalization to new targets.} Our models may perform markedly worse on structurally novel POIs or E3 ligases not represented in training. We explored target-based splits but found they produced highly imbalanced sets, complicating fair evaluation. Developing robust strategies for out-of-distribution generalization remains an open challenge. \textit{Lack of 3D information.} We demonstrated that current best models rely on 1D/2D molecular representations. Incorporating 3D ternary complex structure information, now increasingly available through co-folding tools, could improve predictions by capturing binding pocket complementarity and protein-protein interfaces. This however requires reliable structure prediction for arbitrary POI-PROTAC-E3 combinations, which remains an active research area. \textit{Cell context.} While we encode cell line identity, we do not model the underlying biological differences (\eg, E3 ligase expression levels) that drive cell-specific degradation outcomes. Integrating transcriptomic or proteomic context could improve predictions but requires matched omics data currently unavailable at scale. \textit{Benchmark scope.} TACK focuses on CRBN- and VHL-recruiting PROTACs, which dominate current datasets. Performance on emerging E3 ligases (\eg, IAPs) remains untested and may differ substantially. \textit{Dual-use concerns.} TACK may in theory be misused to predict activity against beneficial targets, with the aim of designing toxic PROTACs. However, the lack of public data on PROTAC toxicity makes this a weak concern at this stage.

\section{Conclusion}

We introduce TACK, a standardized benchmark dataset aggregating 3,514 unique PROTACs with 6,561 degradation endpoints, and show a rigorous statistical evaluation of ML approaches for PROTAC activity prediction. Our findings yield three actionable insights for the field. First, classical tree-based methods outperform DL models on current datasets, suggesting that architectural sophistication should match data availability. Second, potency ($DC_{50}$) is substantially more predictable than efficacy ($D_{max}$), reflecting the greater dependence of maximum degradation on cellular factors not captured by current standard representations. Third, ensemble-based uncertainty quantification provides calibrated confidence estimates that can guide future prioritization in active learning workflows.

These results establish new baseline expectations for PROTAC property prediction and highlight the continued importance of data curation, standardized evaluation protocols, and thoughtful feature engineering. As PROTAC databases expand and structure prediction tools mature, we anticipate that the gap between simple and complex models will narrow---but principled benchmarking will remain essential to distinguish genuine methodological advances from overfitting to narrow chemical series. TACK, our evaluation framework, and all best models reported herein are made publicly available to support reproducible progress in this emerging area of ML-guided drug discovery.




\section*{GenAI Disclosure}
The authors used generative AI for polishing parts of the text and the code, using the following models: Claude Sonnet 4.5, Claude Opus 4.5 and Gemini Pro. All the authors have reviewed and approve of the final version of the submitted manuscript.

\section*{Software and Data}
Code to reproduce all work, including the data curation pipeline, model training, and statistical analysis, can be found in the following repository: \textbf{\href{https://github.com/ribesstefano/TACK/}{\uline{https://github.com/ribesstefano/TACK/}}}

The curated TACK dataset can be inspected and obtained from HuggingFace: \textbf{\href{https://huggingface.co/datasets/ailab-bio/TACK}{\uline{https://huggingface.co/datasets/ailab-bio/TACK}}}

All best models presented herein, including ensemble models, are available via Zenodo: \textbf{\href{https://zenodo.org/records/15691822}{\uline{https://zenodo.org/records/15691822}}}

\begin{acks}
SR and RM acknowledge funding provided by the Chalmers Gender Initiative for Excellence (Genie). RM and ND acknowledge funding provided by the Wallenberg AI, Autonomous Systems, and Software Program (WASP), supported by the Knut and Alice Wallenberg Foundation.
The authors thank Yossra Gharbi, Alexander Persson, and Felix Erngård for helpful discussions.
The computations and data storage were enabled by resources provided by Chalmers e-Commons and by the National Academic Infrastructure for Supercomputing in Sweden (NAISS), partially funded by the Swedish Research Council through grant agreement no. 2022-06725. The authors declare no competing interests.
\end{acks}

\bibliographystyle{ACM-Reference-Format}
\balance
\bibliography{reference}

\appendix

\section{CV Folds Label Distribution}
\label{app:cv_distributions}

\begin{figure*}[h]
    \centering
    \includegraphics[width=\textwidth]{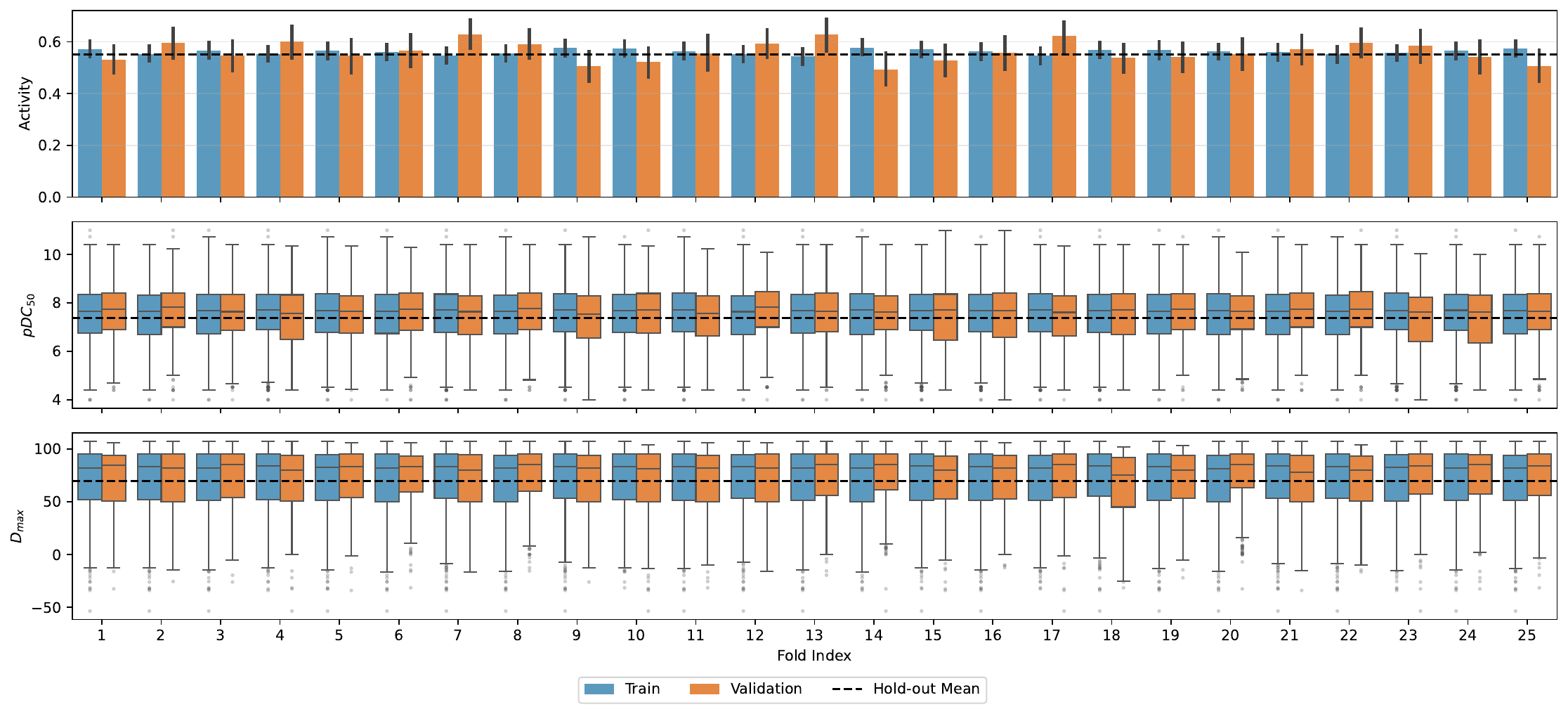}
    \caption{Distribution of the target values across the 5$\times$5 CV folds for each prediction task.}
    \label{fig:cv_distributions}
\end{figure*}

Figure \ref{fig:cv_distributions} shows the distribution of the target labels of the 5$\times$5 CV folds used in each of the three tasks explored in this work, together with the mean of the hold-out set labels. This plot demonstrates that the CV folds are balanced.

\section{Normality Diagnostic \& Effect Size}
\label{app:normality_diagnostic}

\begin{figure}[h!]
    \centering
    \includegraphics[width=\linewidth]{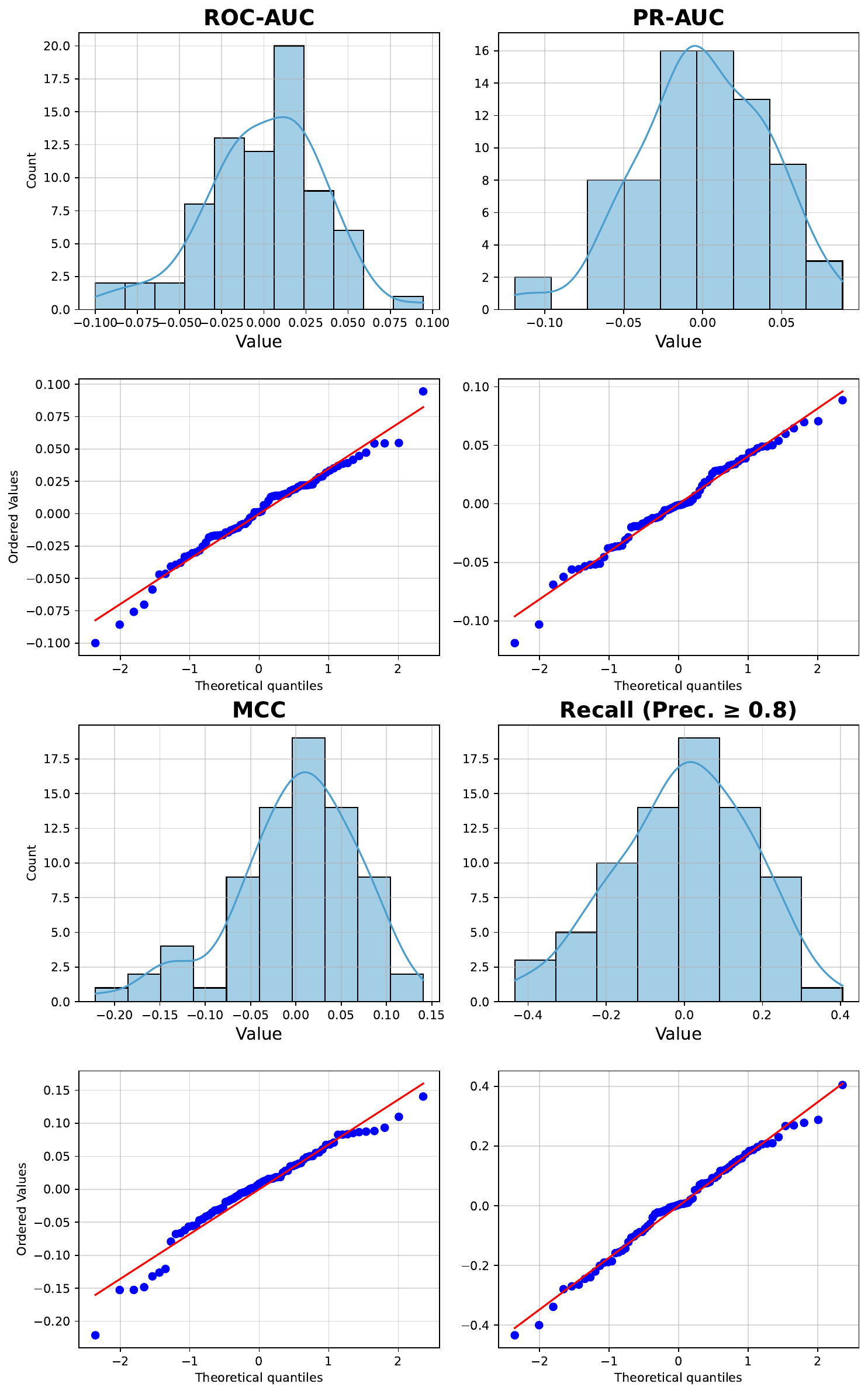}
    \caption{Normality diagnostic of the performance metric values on the fold validation sets of the three models: XGBoost$^{(\ddag)}$, MLP$^{(\ddag)}$, and PROTAC-STAN.}
    \label{fig:normality_diagnostic}
\end{figure}

\begin{figure}[h!]
    \centering
    \includegraphics[width=\linewidth]{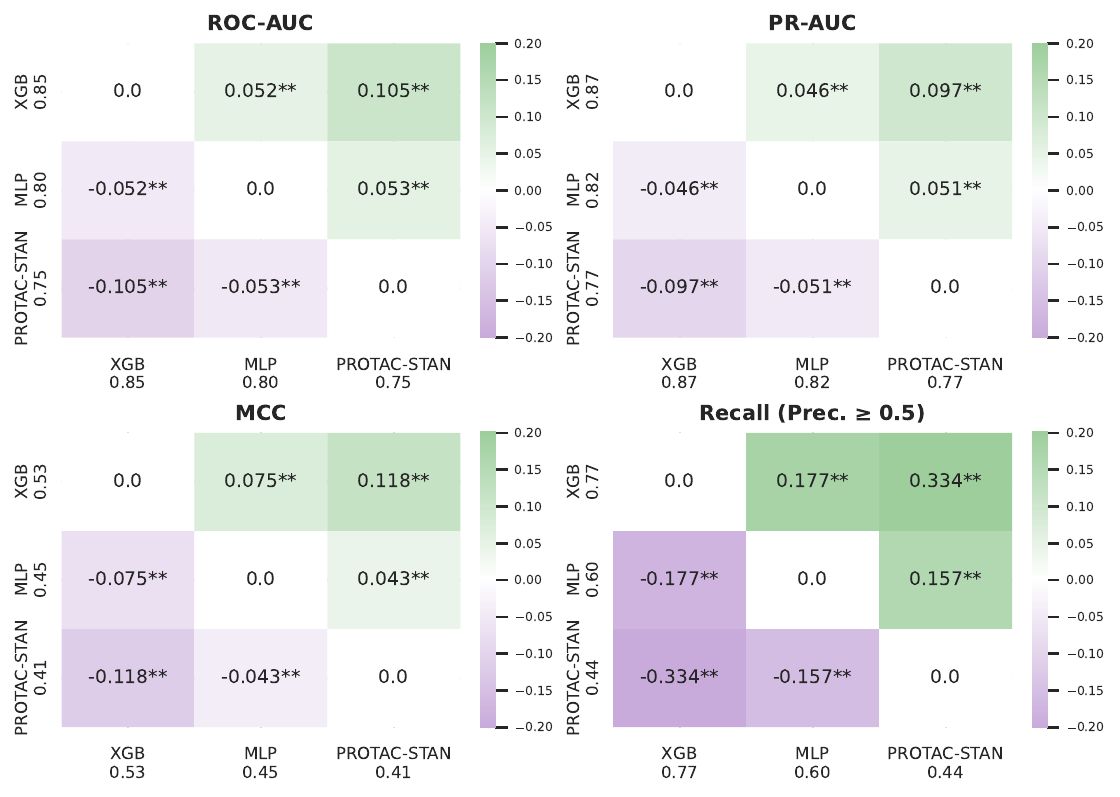}
    \caption{Effect sizes of the performance metric values on the fold validation sets of the three models: XGBoost$^{(\ddag)}$, MLP$^{(\ddag)}$, and PROTAC-STAN. Values marked with $^{**}$ have a significance of: $0.001 \leq p < 0.1$.}
    \label{fig:effect_size_bin_val}
\end{figure}

Figure \ref{fig:normality_diagnostic} visually confirms the assumption of normality distribution of the metrics values of the activity prediction models.
This allowed us to perform parametric Tukey’s HSD test to compare the performance of the three activity prediction models, XGBoost$^{(\ddag)}$, MLP$^{(\ddag)}$, and PROTAC-STAN, as described in Section \ref{sec:model_comparison}.
Detailed effect sizes from performing Tukey's HSD are visualized in Figure \ref{fig:effect_size_bin_val}.
The top-left-most model indicates the best performing model for a given metric.

\section{Model Details}
\label{app:model_details}

We developed a multilayer perceptron (MLP) architecture with flexible depth and width. The MLP consists of a variable number of hidden layers with dimensions sampled from predefined configurations (\eg, \textsc{[256]}, \textsc{[512, 256]}, \textsc{[1024, 512, 256]}), followed by a regression head (\ie, a linear layer) with configurable depth (1--3 layers). Each hidden layer incorporates optional normalization (batch normalization or layer normalization), dropout regularization (0.0--0.5), and activation functions (ReLU, GELU, or SiLU). The model weights are initialized using Kaiming initialization~\cite{he2015delving}, with the first layer initialized from $\mathcal{N}(0, 1/\sqrt{d_{\text{in}}})$ and subsequent layers from $\mathcal{N}(0, \sqrt{2}/\sqrt{d_{\text{in}}})$ to account for ReLU nonlinearities. Training leveraged gradient clipping (0.5--2.0), mixed-precision training (16-bit), and early stopping with a patience of 5 epochs based on the validation loss.

PROTAC-STAN~\cite{chen2025} is a graph neural network architecture designed to predict the degradation activity of PROTACs using a ternary attention network (TAN) layer~\cite{Kim2018}. The model encodes PROTAC molecules using a two-layer edge-aware graph convolutional network (GCN) with global max pooling, while protein sequences (POI and E3 ligase) are represented using ESM-S \cite{zhang2024esms} embeddings processed through a two-layer fully connected adapter. The TAN layer is a specialized multi-head attention mechanism that explicitly models the complex three-way interactions between the POI, PROTAC, and E3 ligase embeddings computing a joint representation via tensor outer products. The joint representation is then fed into a classifier MLP with batch normalization and ReLU activation to compute the probability of degradation activity. For PROTAC-STAN, we trained the model from randomly initialized weights using the hyperparameters reported in the original publication~\cite{chen2025}. Early stopping with patience of 30 epochs was applied for consistency with the other models.

\section{Feature Sets}
\label{app:feature_sets}

Table~\ref{tab:feature_sets} lists the combinations of the different encodings of the features we explored in our work.
We limited our analysis to the 10 most promising feature sets.
For example, we avoided using 1- and 2-grams vectorization of the dozen available E3 ligase amino acid sequences, as they resulted in vectors of dimension $\sim$4K, making the models prone to overfitting.
The column `Assay' reports whether the assay type  (one-hot encoded) and the experiment time (real value) are included in the set.

\begin{table}[h!]
\centering
\caption{The feature set configurations evaluated across tasks. Shorthand for feature sets is defined in Sec. \ref{sec:feature_extraction}.}
\label{tab:feature_sets}
\begin{center}
\begin{small}
\begin{sc}
\resizebox{\linewidth}{!}{%
\begin{tabular}{cccccc}
\toprule
\textbf{Config} & \textbf{Cell} & \textbf{E3} & \textbf{POI} & \textbf{Mol} & \textbf{Assay} \\
\midrule
1 & OneHot & OneHot & OneHot & Desc+FP & \checkmark \\
2 & OneHot & OneHot & OneHot & FP & \checkmark \\
3 & OneHot & OneHot & OneHot & Desc & \checkmark \\
4 & OneHot & OneHot & Vec & Desc & \checkmark \\
5 & Text & ESM-S-PCA & ESM-S-PCA & Desc+FP & \checkmark \\
6 & Text & ESM-S-PCA & ESM-S-PCA & Desc & \checkmark \\
7 & Text & OneHot & Vec & Desc+FP & \checkmark \\
8 & Text & OneHot & OneHot & Desc & \checkmark \\
9 & Text & OneHot & Vec & Desc & \checkmark \\
10 & --- & ESM-S & ESM-S & Desc+FP & --- \\
\bottomrule
\end{tabular}
}%
\end{sc}
\end{small}
\end{center}
\end{table}

\section{Hyperparameters}
\label{app:hyperparameters}

In training the XGBoost and MLP models, we performed hyperparameter optimization via Optuna~\cite{akiba2019optuna}, using 20 and 100 trials for the XGBoost and MLP models, respectively (search space summarized in Table \ref{tab:hyperparams}).
The optimization objective was to minimize the root mean squared error (RMSE) for regression tasks and maximize accuracy for classification.
The hyperparameters leading to the best performance in the validation set of the first fold were selected and used to train the 25 models on all folds.
Finally, at each fold, we set a different random seed for the model weights initialization to introduce variance in the models' parameters.

\subsection{Hyperparameter Space}

Hyperparameters for both XGBoost and MLP models were optimized using the tree-structured Parzen estimator (TPE) algorithm~\cite{bergstra2011algorithms} implemented in Optuna~\cite{akiba2019optuna}, with 20\% of trials allocated for random exploration and multivariate parameter modeling enabled. Table~\ref{tab:hyperparams} summarizes the search space explored during hyperparameter optimization for each model. For XGBoost, we tuned 8 parameters controlling tree complexity and regularization. For MLP models, we optimized 9 architectural and training parameters, including network depth/width, regularization, and learning rate schedules.

\begin{table}[h!]
\centering
\caption{Hyperparameter search spaces for the XGBoost and MLP models.}
\label{tab:hyperparams}
\begin{small}
\begin{sc}
\resizebox{\linewidth}{!}{%
\begin{tabular}{lll}
\toprule
\textbf{Model} & \textbf{Hyperparameter} & \textbf{Search Space} \\
\midrule
\multirow{8}{*}{XGB} 
& Learning rate & $[10^{-3}, 10^{-1}]$ (log-uniform) \\
& Max depth & $[3, 9]$ (integer) \\
& Min child weight & $[1, 25]$ (integer) \\
& Subsample & $[0.4, 1.0]$ (uniform) \\
& Column sample by tree & $[0.5, 1.0]$ (uniform) \\
& L1 regularization ($\alpha$) & $[10^{-3}, 10]$ (log-uniform) \\
& L2 regularization ($\lambda$) & $[10^{-3}, 10]$ (log-uniform) \\
& Gamma & $[10^{-3}, 10]$ (log-uniform) \\
\midrule
\multirow{9}{*}{MLP} 
& Hidden dimensions & Categorical: [256], [512], [256, 128], \\
& & [512, 256], [512, 256, 128], [1024, 512], [1024, 512, 256] \\
& Dropout & $[0.0, 0.5]$ (step 0.1) \\
& Head dropout & $[0.0, 0.3]$ (step 0.1) \\
& Head depth & $[1, 3]$ (integer) \\
& Normalization type & Categorical: None, batch, layer \\
& Activation function & Categorical: ReLU, GELU, SiLU \\
& Learning rate & $[10^{-5}, 10^{-2}]$ (log-uniform) \\
& LR scheduler & Categorical: cosine, reduce-on-plateau \\
& Warmup ratio & $[0.0, 0.2]$ (step 0.05) \\
& Gradient clip value & $[0.5, 2.0]$ (step 0.5) \\
\bottomrule
\end{tabular}
}%
\end{sc}
\end{small}
\end{table}

\begin{table}[h!]
\centering
\caption{Optimized hyperparameters for the best performing models marked with $(\dag)$ and $(\ddag)$ in Table \ref{tab:best_features_configurations}. XGBoost models used 2000 estimators with early stopping after 30 rounds. MLP models were trained for up to 200 epochs with early stopping and mixed precision (16-bit).}
\label{tab:best_model_hyperparams}
\begin{center}
\begin{small}
\begin{sc}
\resizebox{\linewidth}{!}{%
\begin{tabular}{lcc}
\toprule
\multicolumn{3}{c}{\textbf{XGBoost Models}} \\
\midrule
\textbf{Hyperparameter} & \textbf{\Dmax$^{(\dag)}$} & \textbf{Activity$^{(\ddag)}$} \\
\midrule
Learning rate & 0.0159 & 0.0018 \\
Max depth & 7 & 6 \\
Min child weight & 1 & 1 \\
Gamma & 0.005 & 0.120 \\
Subsample & 0.982 & 0.946 \\
Column sample by tree & 0.916 & 0.629 \\
L1 regularization ($\alpha$) & 0.007 & 0.447 \\
L2 regularization ($\lambda$) & 0.005 & 0.018 \\
\midrule
\multicolumn{3}{c}{\textbf{MLP Models}} \\
\midrule
\textbf{Hyperparameter} & \textbf{\pDC$^{(\dag)}$} & \textbf{Activity$^{(\ddag)}$} \\
\midrule
Hidden dimensions & [256] & [512, 256, 128] \\
Dropout & 0.0 & 0.0 \\
Head depth & 1 & 1 \\
Head dropout & 0.1 & 0.1 \\
Normalization & Layer & Layer \\
Learning rate & 0.00075 & 0.00036 \\
LR scheduler & Reduce-on-plateau & Reduce-on-plateau \\
Warmup ratio & 0.0 & 0.05 \\
Activation & ReLU & SiLU \\
Gradient clip value & 1.0 & 1.5 \\
\bottomrule
\end{tabular}
}%
\end{sc}
\end{small}
\end{center}
\end{table}

\subsection{Best Models Hyperparameters}

Table~\ref{tab:best_model_hyperparams} lists the hyperparameters of the best models reported in Table \ref{tab:best_features_configurations}, specifically those marked with ${\dag}$ and ${\ddag}$.

\end{document}